\documentclass[conference]{IEEEtran}
\IEEEoverridecommandlockouts

\usepackage{cite}
\usepackage{amsmath,amssymb,amsfonts}
\usepackage{algorithmic}
\usepackage{graphicx}
\usepackage{textcomp}
\usepackage{xcolor}
\usepackage{balance}

\usepackage{hyperref}
\usepackage{url}
\usepackage{xspace}
\usepackage{multirow}
\usepackage{booktabs}
\usepackage{threeparttable}
\usepackage{array}
\usepackage{tcolorbox}
\usepackage{inconsolata}
\usepackage{silence}
\usepackage{caption}
\usepackage{subcaption}

\def\BibTeX{{\rm B\kern-.05em{\sc i\kern-.025em b}\kern-.08em
    T\kern-.1667em\lower.7ex\hbox{E}\kern-.125emX}}
\begin{document}

\newcommand{\ourname}{\textsc{ApiRAT}\xspace} 
\newcommand{\apidata}{\textsc{ApiSeqData}\xspace} 
\newcounter{prompt}
\renewcommand{\theprompt}{Prompt~\arabic{prompt}}
\definecolor{mygray}{RGB}{128,128,128}

\definecolor{darkred}{rgb}{0.8, 0.1, 0.1}
\definecolor{darkgreen}{rgb}{0.1, 0.6, 0.1}
\definecolor{lightgray}{gray}{0.9}
\definecolor{darkgray2}{RGB}{90,90,90}
\definecolor{codegreen}{RGB}{56,157,60}
\definecolor{codeblue}{RGB}{19,119,199}
\definecolor{codepurple}{RGB}{162,65,163}
\definecolor{codeorg}{RGB}{240,136,26}
\definecolor{codegreen2}{RGB}{0,193,6}

\newcommand{\todoc}[2]{{\textcolor{#1}{\textbf{[#2]}}}}
\newcommand{\todoblue}[1]{\todoc{blue}{\textbf{#1}}}
\newcommand{\todored}[1]{\todoc{red}{\textbf{#1}}}
\newcommand{\tododarkred}[1]{\todoc{darkred}{\textbf{#1}}}

\newcommand{\wang}[1]{\todoblue{wang: #1}}
\newcommand{\gu}[1]{\todored{gu: #1}}
\newcommand{\shen}[1]{\tododarkred{shen: #1}}

\title{\fontsize{23.8pt}{30pt}\selectfont \ourname: Integrating Multi-source API Knowledge for Enhanced Code Translation with LLMs}

\author{\IEEEauthorblockN{Chaofan Wang, Guanjie Qiu, Xiaodong Gu, Beijun~Shen$^{\dagger}$}\\
\IEEEauthorblockA{\textit{School of Computer Science, Shanghai Jiao Tong University, Shanghai, China}\\
\{chaofwang, qiuguanjie, xiaodong.gu, bjshen\}@sjtu.edu.cn}}

\maketitle

\begin{abstract}
Code translation is an essential task in software migration, multilingual development, and system refactoring. Recent advancements in large language models (LLMs) have demonstrated significant potential in this task. However, prior studies have highlighted that LLMs often struggle with domain-specific code, particularly in resolving cross-lingual API mappings. 
To tackle this challenge, 
we propose \ourname, a novel code translation method that integrates multi-source API knowledge. \ourname employs three API knowledge augmentation techniques, including API sequence retrieval, API sequence back-translation, and API mapping, to guide LLMs to translating code, ensuring both the correct structure of API sequences and the accurate usage of individual APIs.
Extensive experiments on two public datasets, CodeNet and AVATAR, indicate that \ourname significantly surpasses existing LLM-based methods, achieving improvements in computational accuracy ranging from 4\% to 15.1\%. 
Additionally, our evaluation across different LLMs showcases the generalizability of \ourname. An ablation study further confirms the individual contributions of each API knowledge component, underscoring the effectiveness of our approach.

\end{abstract}


\section{Introduction}
$\let\thefootnote\relax\footnotetext{
$\dagger$ Beijun Shen is the corresponding author.}$

Code translation, namely, migrating source code developed in one programming language into another \cite{definition}, has been a crucial task in modern software development. 
Automatic code translation enables cross-platform compatibility, facilitating the reuse of legacy systems and multilingual programming environments\cite{intertrans}, hence significantly reducing developers' effort in software maintenance \cite{program}.
Recently, large language models (LLMs) such as GPT-4 have demonstrated impressive performance in code translation~\cite{codegeex, gtranseval, stelocoder, selfdebug, exp, spectra}. By pre-training on vast datasets, LLMs learn the mappings between source and target languages, enabling efficient cross-language code conversion. 

\begin{figure}[t!]
    \centering
    \begin{subfigure}[b]{\linewidth}
        \centering
        \includegraphics[width=\linewidth]{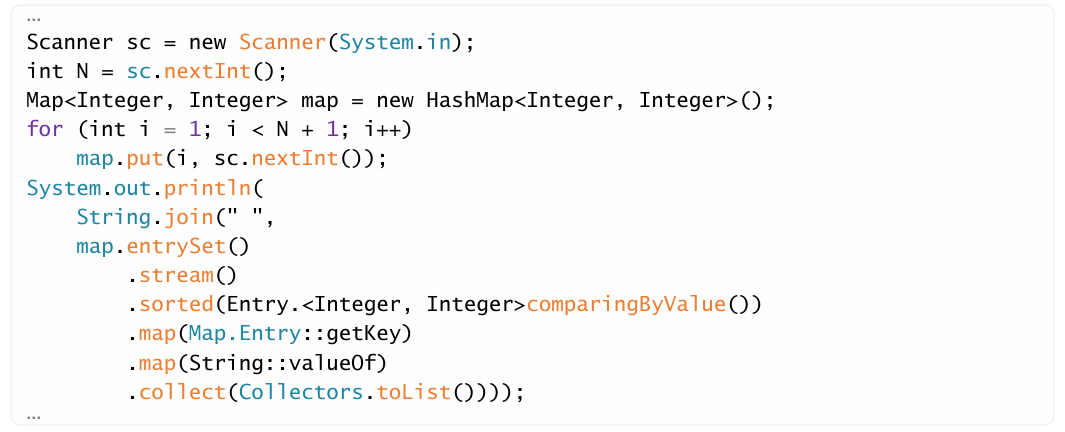}
        \caption{Java Source Code}
        \label{sourcecode} 
    \end{subfigure}
    
    \begin{subfigure}[b]{\linewidth}
        \centering
        \includegraphics[width=\linewidth]{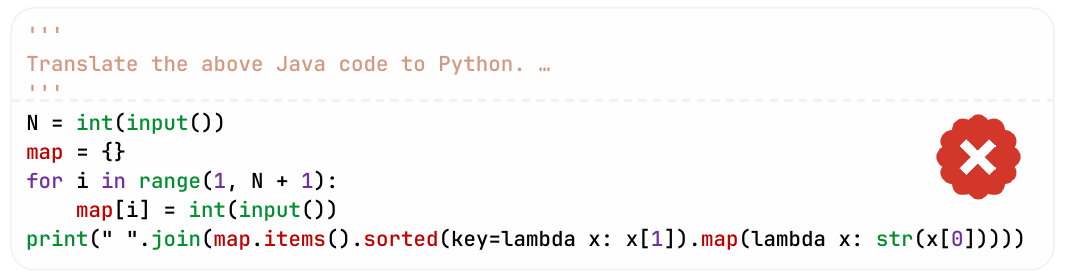}
        \caption{Direct Code Translation}
        \label{direct} 
    \end{subfigure}

    \begin{subfigure}[b]{\linewidth}
        \centering
        \includegraphics[width=\linewidth]{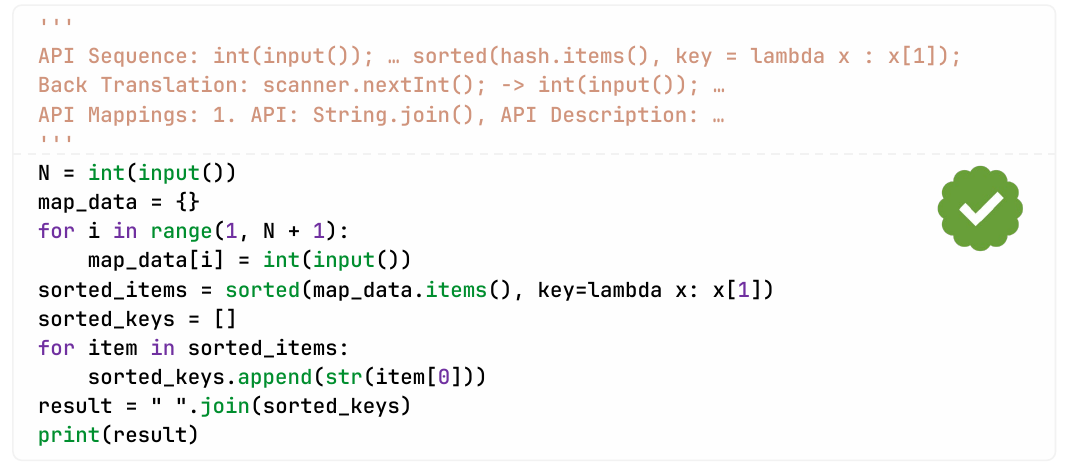}
        \caption{With API Knowledge Augmentation (ours)}
        \label{ours}         
    \end{subfigure}    

    \caption{An Example of Code Translation with LLMs} 
    \label{fig:motivation}    
    \vspace{-7pt}
\end{figure}

Although promising, prior studies\cite{gtranseval, domcoder} have revealed that LLMs are hindered by a deficiency in specialized domain knowledge, exacerbated by the substantial disparities in domain APIs across different programming languages. 
These limitations lead to poor performance of existing LLMs in translating cross-language API calls\cite{lostintranslation}. Specifically, when the target language does not provide a direct equivalent for an API from the source language, LLMs frequently encounter difficulties in generating functionally equivalent code. 
Fig. \ref{fig:motivation} illustrates a Java$\rightarrow$Python translation example for a sorting problem, where direct translation using the LLM results in \texttt{map} API errors stemming from naming conflicts, incorrect method attribution, and output mismatches.

To delve into these challenges, we conduct a preliminary study on API errors in code translation. By analyzing LLM translated code on the CodeNet and AVATAR datasets, we identify that over 60\% of errors are attributed to API mistranslation. We further classify these errors into twelve fault patterns, e.g., misused API calls, missing API calls, improper parameter constraints, and invocation of non-existent APIs. These patterns can be broadly categorized into two types: single API errors and API sequence errors.

To tackle the above mentioned issues, we propose \ourname (\underline{API} \underline{R}etrieval \underline{A}ugmented \underline{T}ranslation), a novel method by integrating external API knowledge into LLM code translation. \ourname retrieves multiple sources of API knowledge and re-translating test-failed programs by taking them as augmentation.
We design three API knowledge retrieval techniques: \textit{API sequence retrieval}, \textit{API sequence back-translation}, and \textit{API mapping}. 
Specifically, we construct an embedding database of API sequences extracted from open-source projects in target languages, enabling the retrieval of target API sequences based on their source counterparts. These retrieved API sequences are then back-translated into the source language, collectively enhancing LLMs to mitigate API sequence errors. Furthermore, we manually curate a high-quality API mapping pool from API documentation with the assistance of LLMs, which facilitates the correction of single API errors through retrieval augmentation.
As illustrated in Fig. \ref{ours}, when augmented with API knowledge recommended by \ourname, the LLM significantly improves its accuracy in cross-language API translations.

We evaluate \ourname on two public datasets, CodeNet and AVATAR, and compare its performance with existing state-of-the-art LLM-based translation methods. The results indicate that \ourname consistently outperforms baseline methods in terms of computing accuracy. Additionally, evaluations across different LLMs demonstrate the generalizability of \ourname. Ablation studies further highlight the critical contribution of each API knowledge for achieving optimal performance.

In summary, the contributions of this paper are as follows:

\begin{itemize}
\setlength\itemsep{0em}
\item \textbf{Empirical Study on API Mistranslation}. We conduct an empirical investigation into API mistranslation errors in code translation, identifying twelve fault patterns. This study provides valuable insights for researchers to better understand these errors and develop improved code translation methods.

\item \textbf{API Knowledge Augmentation Framework}. We propose three novel API knowledge retrieval augmentation techniques, namely \textit{API sequence retrieval}, \textit{API sequence back-translation}, and \textit{API mapping}, to enhance the accuracy and robustness of LLM-based code translation.

\item \textbf{Open-Source Tool and Evaluation}. We have released
our tool on Github\footnote{\url{https://github.com/CodeTransFusion/ApiRAT}} and evaluated its performance on two public benchmarks. The results demonstrate that \ourname significantly surpasses existing LLM-based methods, achieving improvements in computational accuracy ranging from 4\% to 15.1\%.
\end{itemize}

\section{Preliminary Study}
\label{sec:empiricalStudy}
We conduct an empirical study to analyze API mistranslation errors in LLM-based code translation, aiming to identify common fault patterns.

\subsection{Setting}

\textbf{Studied LLMs.} We investigate three state-of-the-art LLMs: StarCoder~\cite{starcoder}, GPT-Turbo-3.5 and GPT-4o-mini~\cite{gpt3}. These models are widely recognized for their capabilities in code translation tasks.

\textbf{Metrics.} We employ Computational Accuracy (CA) \cite{transcoder} as the primary evaluation metric for code translation. CA measures whether the source program and the translated program produce identical outputs given the same inputs. For consistency, only the first generated result (i.e., pass@1) is considered during evaluation.

\textbf{Datasets.} The study is conducted on two widely used public benchmark datasets: CodeNet\cite{codenet} and AVATAR\cite{avatar}. CodeNet comprises 200 Python and 200 Java samples, each accompanied by corresponding test cases. AVATAR, a more challenging algorithmic benchmark, includes 250 Python and 249 Java samples, along with stricter test cases. 

\subsection{Results}

\begin{table}[t!]
    \centering
    \caption{Performance of LLMs for Code Translation}
    \renewcommand{\tablename}{Table}
    \label{tab:empirical}
    \resizebox{\columnwidth}{!}{%
    \begin{tabular}{lcccc}
        \toprule
        \multirow{2}{*}{\textbf{Model}} & \multicolumn{2}{c}{\textbf{CodeNet}} & \multicolumn{2}{c}{\textbf{AVATAR}} \\
        \cmidrule(lr){2-3} \cmidrule(lr){4-5}
         & \textbf{Py-to-Ja} & \textbf{Ja-to-Py} & \textbf{Py-to-Ja} & \textbf{Ja-to-Py} \\
        \midrule
        StarCoder & 48.5\% & 22.0\% & 22.0\% & 20.9\% \\
        GPT-3.5-Turbo & 82.0\% & 63.5\% & 49.6\% & 67.5\% \\
        GPT-4o-mini & 92.0\% & 64.0\% & 63.2\% & 64.7\% \\
        \bottomrule
    \end{tabular} 
    }    
    \begin{tablenotes}
    \item * Py: Python; Ja: Java
    \end{tablenotes}
    \vspace{-5pt}
\end{table}

We investigate the performance of LLMs on code translation tasks. As presented in Table \ref{tab:empirical}, LLMs exhibit significant potential in code translation, highlighting the need for further research. Among them, GPT-4o-mini and GPT-3.5-Turbo demonstrates superior performance compared to the open-source model StarCoder.


\begin{figure*}[hbtp!]
\centerline{\includegraphics[width=0.7\textwidth, trim=0 50 0 50, clip]{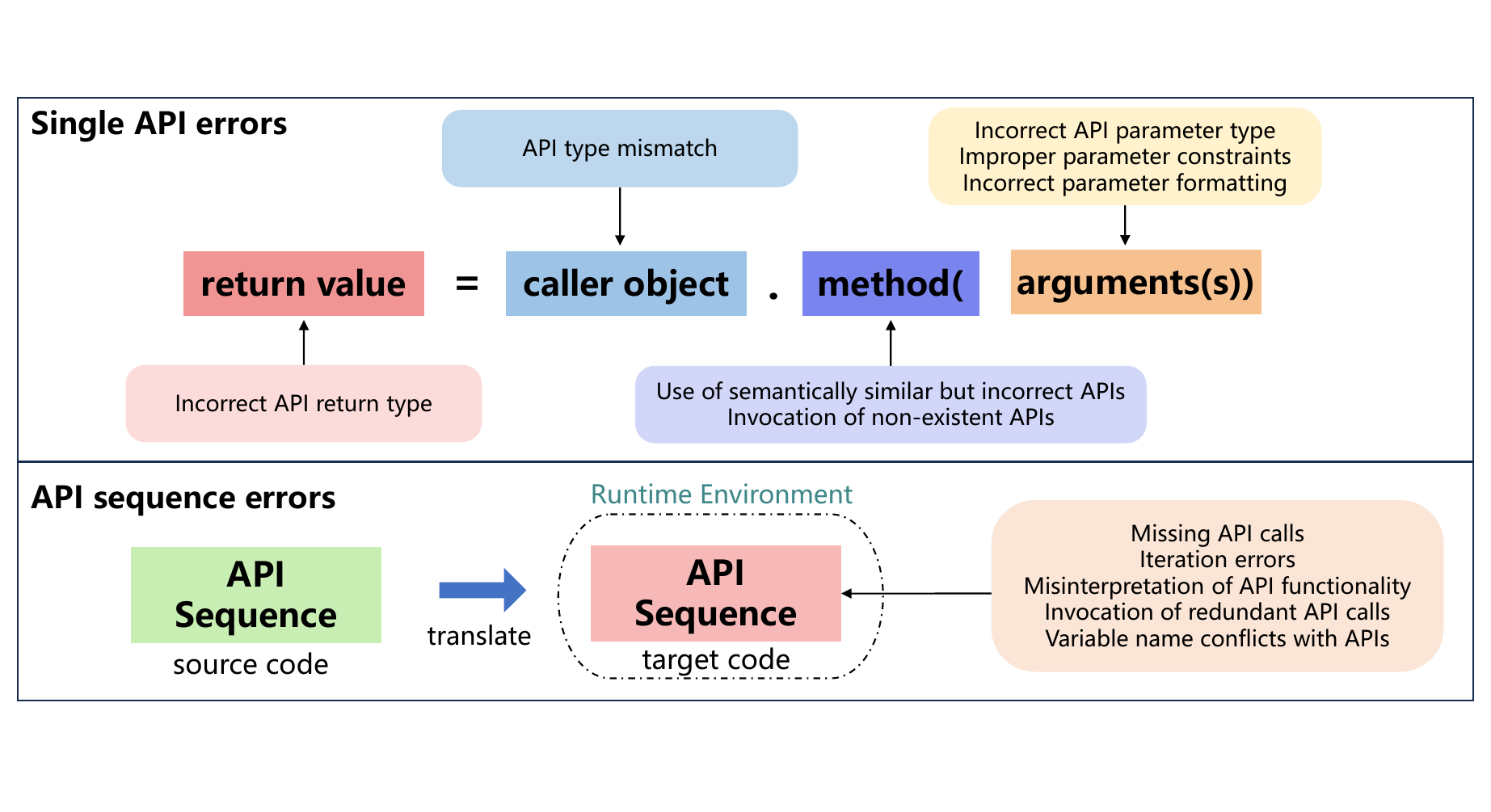}}
    \caption{API Fault Patterns in Code Translation} 
    \label{fig:APIerrors}
    \vspace{-5pt}
\end{figure*}

Through a fine-grained analysis of incorrect translations, we observe that over 60\% of errors originate from API mistranslations. This reveals significant limitations in LLMs' ability to comprehend the nuanced semantics and context-specific requirements of APIs across programming languages. Based on these findings, we further identify twelve API-related fault patterns, which are categorized into single API errors and API sequence errors, as illustrated in Figure \ref{fig:APIerrors}.

Single API errors encompass seven fault patterns:
\begin{itemize}
\item \textbf{Semantically similar but incorrect APIs}: LLMs select a target language API that appears functionally similar to the source API but exhibits subtle semantic differences. For example, mapping Python’s \texttt{join()} to Java’s \texttt{ArrayList$<$Character$>$.valueOf()} introduces unintended delimiters in the output string.

\item \textbf{API type mismatch}: LLMs apply an API to an incompatible object type, causing runtime errors, such as invoking a string method on a char type.

\item \textbf{Incorrect API parameter type}: The API is called with an argument of an invalid type, violating its signature, e.g., passing a string to an integer parameter.

\item \textbf{Incorrect API return type}: LLMs mishandle the API’s return type, resulting in type mismatches, such as assigning a String return value to a char variable.

\item \textbf{Improper parameter constraints}: LLMs generate API calls with arguments that violate required constraints, e.g., providing arguments in ascending order when descending is expected.

\item \textbf{Incorrect parameter formatting}: The argument format does not conform to the API’s requirements, such as providing malformed JSON input.

\item \textbf{Invocation of non-existent APIs}: LLMs generate calls to APIs that do not exist in the target language, e.g., using \texttt{** (count - 1)} for exponentiation in Java, where \texttt{**} is not a valid operator.

\end{itemize}

API sequence errors include five fault patterns:
\begin{itemize}
\item \textbf{Missing API calls}: The translated code omits necessary API calls present in the source language, resulting in incomplete functionality.

\item \textbf{Redundant API calls}: The translated code includes unnecessary API calls absent in the source language, introducing computational overhead.

\item \textbf{Misinterpretation of API functionality}: LLMs misinterpret an API’s behavior, leading to incorrect translations, e.g., failing to recognize that Python’s \texttt{' '.join(...)} inserts spaces between elements.

\item \textbf{Iteration errors}: LLMs incorrectly translate loop structures involving API calls, causing logical inconsistencies, such as misinterpreting boundary conditions in Python$\rightarrow$Java loop conversions.

\item \textbf{Variable name conflicts with APIs}: The translated code introduces variable names that clash with existing API or function names, causing compilation errors due to ambiguity.

\end{itemize}

These findings underscore the limitations of current LLMs in managing API transformations across languages with divergent paradigms, emphasizing the need for novel techniques to address these challenges and improve translation accuracy.

\section{Approach}
\label{sec:approach}


\subsection{Overview}

The code translation task can be formally defined as follows: Given a program \( x = x_{1}, \ldots, x_{p} \) and  a set of test cases \( T_x = \{t_1, \ldots, t_n\} \) in a source language, the objective is to generate a semantically equivalent target program \( y = y_{1}, \ldots, y_{q} \) in the target language. The API sequences in the source and target programs are represented as \( API_x = \{a_1, a_2, \ldots, a_m\} \) and \( API_y = \{a'_1, a'_2, \ldots, a'_t\} \), respectively, where each \( a_i \) and \( a'_i \) corresponds to an API invocation. 

\begin{figure*}[hbtp!]
\centerline{\includegraphics[width=0.9\textwidth, trim=0 0 0 0, clip]{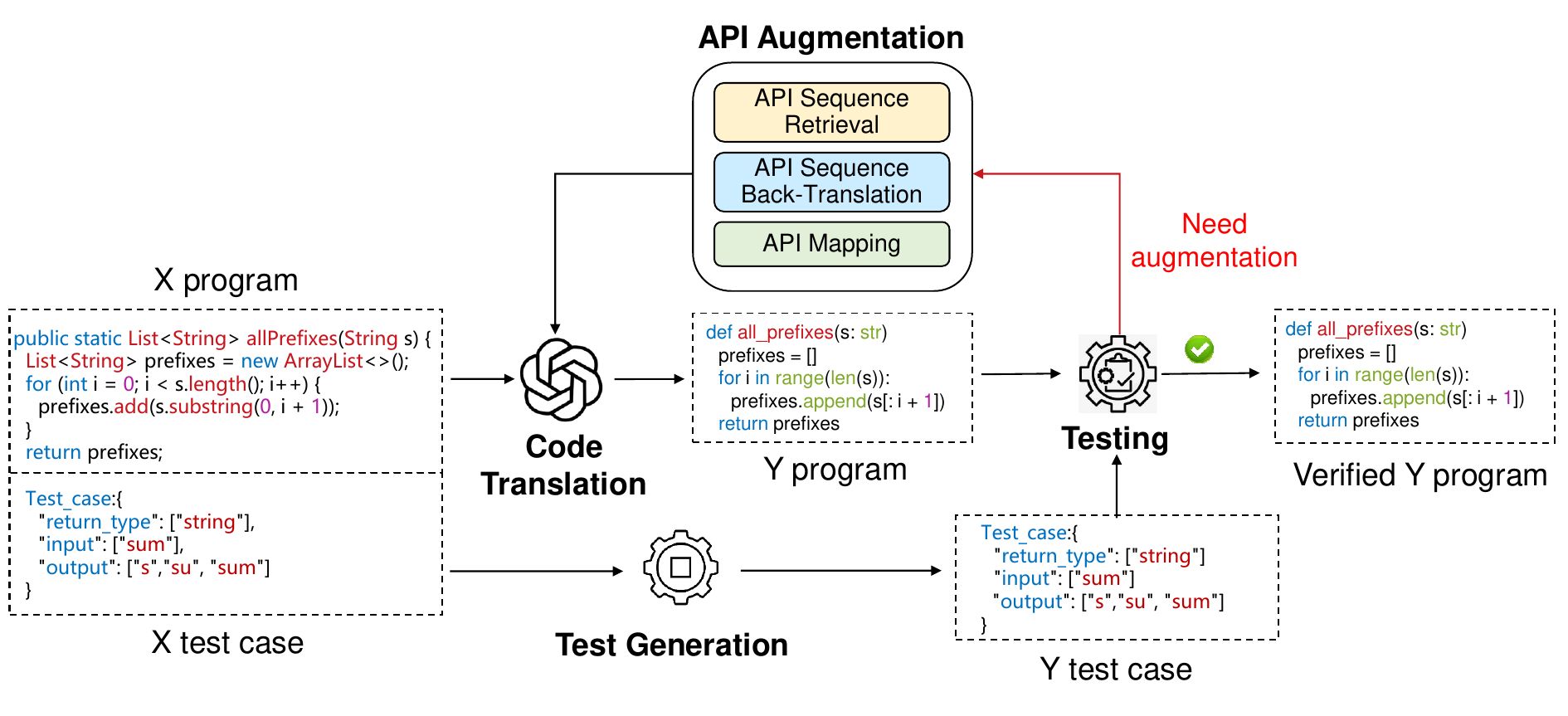}}
    \caption{Overview of \ourname} 
    \label{fig:overview}
    \vspace{-5pt}
\end{figure*}

To address the API mistranslation challenge in this task, we propose \ourname, a novel method that enhances LLMs by integrating external API knowledge. \ourname simultaneously translates both the source program and its corresponding test cases into the target language, enabling self-testing to ensure translation quality. For instances where the initial translation fails, \ourname re-translates the program by augmenting it with API knowledge retrieved from multiple sources, including \textit{API sequence retrieval}, \textit{API sequence back-translation}, and \textit{API mapping}. 

Figure~\ref{fig:overview} presents the framework of \ourname, which consists of the following key steps. First, it instructs LLMs to generate an initial code translation using basic prompts, while simultaneously employing a rule-based test tool to generate the corresponding test cases (Section~\ref{initialTrans}). Second, the translated program is validated against the generated test cases. For instances where the translation fails, \ourname retrieves API knowledge from multiple external sources (Section~\ref{retrieval}, ~\ref{back-translation}, and ~\ref{apiMapping}). Finally, the retrieved API knowledge is integrated into the prompt to guide the final code translation (Section~\ref{KAT}). Each of these key steps is detailed in the subsequent subsections.

\subsection{Initial Translation}
\label{initialTrans}

Initial code translation consists of two key components: direct code translation and test generation.

\textbf{Direct Code Translation}. 
We utilize LLMs to directly generate target program (i.e., the initial target program \(y\)) from the provided source program \(x\). Adopting the ``vanilla-prompting'' translation template \cite{lostintranslation}, each prompt includes four essential elements: (1) the source programming language (\texttt{\$SOURCE\_LANG}), (2) the target programming language (\texttt{\$TARGET\_LANG}), (3) the source program to be translated (\texttt{\$TARGET\_LANG}), and (4) a natural language instruction for code translation, as shown in Prompt 1. 

\begin{tcolorbox}[colframe=darkgray2, colback=white, coltitle=white, colbacktitle=darkgray2, 
title=Prompt 1: Prompt for Direct Code Translation,
boxrule=0.8pt,
fonttitle=\mdseries\small, fontupper=\ttfamily\footnotesize, rounded corners]
\label{prompt1}
\textcolor{mygray}{\#\#\# Unformatted source code} \\
\textcolor{codeblue}{\$SOURCE\_CODE} \\ 
\\
Translate the above \textcolor{codepurple}{\$SOURCE\_LANG} code to \textcolor{red}{\$TARGET\_LANG}. Print only the \textcolor{red}{\$TARGET\_LANG} code, end with comment \textcolor{codegreen}{"|End-of-Code|"}.
\end{tcolorbox}

\textbf{Test Generation}. 
In parallel, we employ a rule-based test generation tool released by Tao et al. \cite{polyhumaneval} to produce equivalent test cases in the target language, ensuring compatibility with the target environment and enabling error-free execution. This tool automatically generates test programs across multiple programming languages using problem-specific metadata. Following the tool's specifications, we manually define metadata for each problem, including a function signature with standardized data types consistently applied across languages, as well as corresponding test cases that specify inputs and expected outputs.
Consequently, \( T_y = \{t'_1, t'_2, \ldots, t'_n\} \) is generated to verify that \( y \) preserves semantic equivalence to \( x \).

\subsection{Retrieving API Sequences}
\label{retrieval}

Given the source API sequence \(API_x = \{a_1, a_2, \ldots, a_m\}\), \ourname retrieves the top-k most similar target API sequences \( API_y = \{a'_1, a'_2, \ldots, a'_t\} \) from the vector database for API sequences in the target language. 

In the offline stage, we construct two vector databases for API sequences in Java and Python by leveraging GitHub repositories. Specifically, we first rank Java and Python projects on GitHub by popularity (denoted by the stars) and select the top 300 most popular projects. From these projects, we retrieve source files written in Python 3 or Java 11 and segment them into code snippets. Each code snippet represents a function, with comments, blank lines, and redundant spaces removed. We then utilize Tree-sitter\footnote{\url{https://github.com/tree-sitter/tree-sitter}} to parse the code snippets and construct their abstract syntax trees (ASTs). Employing a depth-first traversal strategy, we identify API call names and their associated parameters by examining node types and attributes of ASTs. We then randomly select 200,000 non-redundant API sequences per language and subsequently transform these sequences into high-dimensional embeddings using the text-embedding-3-large\footnote{\url{https://platform.openai.com/docs/guides/embeddings}} model. Consequently, the vector databases for Java and Python each contain 200,000 records, structured in the format \textless Target API Sequence, Embedding Vector\textgreater.

In the online stage, taking a program in the source language as input, \ourname extracts its API sequence by syntax parsing and encodes the sequence into an embedding using text-embedding-3-large. \ourname then identifies relevant target API sequences by calculating the cosine similarity between the current embedding and the embeddings stored in the database, subsequently retrieving the top-k most similar samples.

\subsection{Back-Translating API Sequences}
\label{back-translation}

After retrieving target API sequences, we back-translate them into their corresponding source API sequences. This back-translation strategy enables the model to concentrate on the core logic of the source program, thereby improving translation accuracy. Furthermore, by comparing the back-translated API sequences with the original source API sequences, the model can detect and learn from incorrect API mappings. This comparative analysis enhances the model's understanding of API context and reduces translation errors.

Specifically, for each target API sequence \( API_y=\{a'_1,\ldots,a'_t\}\) in the retrieved top-K list, we leverage the LLM to generate a semantically equivalent source API sequence \( API_{\hat{x}} = \{\hat{a}_1, \ldots, \hat{a}_t\} \).
To ensure precise correspondence between API sequences, we design a structured prompt (i.e., Prompt 2) to obtain a detailed mapping from target API sequences to their source equivalents, serving as a reliable reference for subsequent translation. This process effectively establishes a mapping function from target to source API sequences.

\begin{tcolorbox}[colframe=darkgray2, colback=white, coltitle=white, colbacktitle=darkgray2, 
title=Prompt 2: Prompt for Back-Translation, 
boxrule=0.8pt,
fonttitle=\mdseries\small, fontupper=\ttfamily\footnotesize, , rounded corners]
\refstepcounter{prompt}
\label{prompt2}
Please provide a detailed mapping of \textcolor{red}{\$TARGET\_LANG} API sequence to their \textcolor{codeblue}{\$SOURCE\_CODE} equivalents. \\
\\
\textcolor{mygray}{\#\#\# Target Api sequence} \\
\textcolor{magenta}{\$TARGET\_APIS} 
\end{tcolorbox}

\subsection{API Mapping}
\label{apiMapping}

While API sequence retrieval and back-translation strategies address the mistranslation of API sequences, the API mapping strategy specifically targets single API errors. It not only identifies equivalent APIs in the target language for a given source API but also provides detailed descriptions of usage discrepancies, thereby enriching LLMs with fine-grained API knowledge.

In the offline stage, we construct high-quality API mapping sample pools \( S = \{s_1, \ldots, s_e\} \), where each \( s_i \) represents an \( X\rightarrow Y \) mapping between a source language API and its corresponding target language API(s).
Specifically, we extract APIs from official documentation (e.g.,  The Python Standard Library Reference Manual \footnote{\url{https://docs.python.org/3.10/library/index.html}}, Java API Specification \footnote{\url{https://docs.oracle.com/en/java/javase/11/docs/api/index.html}}), manually deduplicating redundant examples. 
Using an LLM, we translate source APIs into target equivalents, explicitly documenting differences and usage restrictions. Each mapping is manually reviewed and validated against the target language's official documentation. Problematic mappings are revised, and ambiguous cases are resolved by writing and testing additional code for accuracy. The API mappings are vectorized using the text-embedding-3-large model and stored in a vector database. 
The resulting API mapping pools for Java$\rightarrow$Python and Python$\rightarrow$Java contain 586 and 179 records, respectively. Each record includes: (1) the source API and its vector representation, (2) equivalent target APIs, (3) a functional description, and (4) usage limitations and discrepancies to prevent inconsistencies.

In the online stage, \ourname retrieves API mappings from the pool \( S \) for each individual API in the source API sequence \( API_x = \{a_1, \ldots, a_m\} \). 
For a given \( a_i \), the corresponding mappings are retrieved from \( S \) by calculating the cosine similarity between \( a_i \) and the source API in each mapping pair. 
The top-N most similar API mappings are selected, and a \( unique \) operation is applied to remove duplicate entries.

\subsection{Knowledge-Augmented Translation}
\label{KAT}

Finally, with the retrieved API knowledge from previous steps, we design an knowledge-augmented prompt to refine the initial translation, as shown in Prompt 3. It not only leverages the API insights but also incorporates the back-translation history to ensure continuity and contextual relevance. Moreover, the prompt includes the user-provided source program, along with the corresponding API sequences and mappings, which collectively guide the model in translating the program from the source to the target programming language. 

\begin{tcolorbox}[
    colframe=darkgray2, colback=white, coltitle=white, colbacktitle=darkgray2, 
    squeezed title=Prompt 3: Prompt for Knowledge-Augmented Translation, 
    boxrule=0.8pt,
    fonttitle=\mdseries\small, fontupper=\ttfamily\footnotesize, rounded corners
]
\refstepcounter{prompt}
\label{prompt3}
\textcolor{gray}{\#\#\# Previous Conversation (Prompt 1)} \\
\textcolor{codegreen2}{\$CONVERSATION\_HISTORY} \\

\textcolor{gray}{\#\#\# Source to Target API Mappings \\
\textcolor{codeorg}{\$API\_MAPPINGS}} \\

Below is the input source code written in \textcolor{codepurple}{\$SOURCE\_LANG} that you should re-write into \textcolor{red}{\$TARGET\_LANG} programming language. Use the \textcolor{codepurple}{\$SOURCE\_LANG} to \textcolor{red}{\$TARGET\_LANG} API Mappings and API Sequence above as references. Give me only the translated \textcolor{red}{\$TARGET\_LANG} code. Do not add explanations, comments, annotations, or anything else. \\

\textcolor{gray}{\#\#\# Unformatted Source Code} \\
\textcolor{codeblue}{\$SOURCE\_CODE}

\end{tcolorbox}

If the translated program passes the final tests, the enhancement is considered effective; otherwise, the process terminates, indicating that the translation remains incorrect.

\section{Evaluation}
\label{sec:evaluation}

We evaluate \ourname by addressing the following research questions:

\emph{\textbf{RQ1}. How effective is \ourname in code translation?}
We compare \ourname against state-of-the-art approaches to assess its effectiveness in code translation tasks.

\emph{\textbf{RQ2}. How does \ourname perform across different backbone LLMs?}
To investigate the impact of different LLMs, we replace the backbone with widely-used LLMs and analyze their performance variations.

\emph{\textbf{RQ3}. What is the contribution of each API knowledge component in \ourname?}
We evaluate the impact of \textit{API sequence retrieval}, \textit{API sequence back-translation}, and \textit{API mapping} on \ourname's performance by incrementally integrating these knowledge components.

\emph{\textbf{RQ4}. How effective is the API knowledge retriever in \ourname?}
We construct a benchmark dataset for the API knowledge retrieval task and compare the performance of our retriever against several state-of-the-art text retrieval algorithms and code search models.

\emph{\textbf{RQ5}. How do parameter configurations affect the translation performance?}
We conduct ablation studies on two critical parameters, namely the number of API mapping pairs and the number of candidate API sequences, to analyze their impact on code translation performance.

\subsection{Experimental Setup}

\textbf{Baselines.}
We compare \ourname with direct LLM-based translation\cite{lostintranslation} and the following three state-of-the-art retrieval-augmented methods:

\begin{itemize}
    \item \textbf{LLM Two-Step Translation}: This method adopts an ``APIs-then-code'' strategy. It extracts the API call sequence from the source program and employs an LLM to translate the sequence into the target language. The source program, along with the translated API sequence, is then used to prompt the LLM to generate the target program.
 
    \item \textbf{EXP} \cite{exp}: This method generates an explanation of the source program prior to translation. Explanations can vary in granularity, ranging from brief summaries to detailed line-by-line descriptions. Since the translation template of EXP requires additional information such as class declarations and target method signatures, we substitute these elements with placeholders such as ``$\{\{\{ToFill\}\}\}$''.

    \item \textbf{SpecTra} \cite{spectra}: This method leverages multimodal specifications to enhance code translation. It generates three types of specifications: (1) static specifications, including input/output formats and function pre/post-conditions, (2) input-output specifications, and (3) descriptive specifications, which are then used to guide the translation process.
    
\end{itemize}

\textbf{Metrics.} We adopt Computational Accuracy (CA)\cite{transcoder} as the evaluation metric for code translation and Precision@1 for API sequence retrieval. 

\begin{table}[t!]
    \caption{Statistics of the Evaluation Datasets} 
    \label{tab:datasets}
    \centering
    \resizebox{\columnwidth}{!}{%
    \begin{tabular}{c c c c c c}
    \toprule
    \textbf{Dataset} & \textbf{Task} & \textbf{Source} & \textbf{Target} & \textbf{\# samples} & \textbf{\# test cases} \\
    \midrule
    \multirow{2}{*}{CodeNet} & \multirow{2}{*}{Translation} & Java & Python & 200 & 200 \\
    &  & Python & Java & 200 & 200 \\
    \midrule
    \multirow{2}{*}{AVATAR} & \multirow{2}{*}{Translation} & Java & Python & 249 & 6,255 \\
    &   & Python & Java & 250 & 6,255 \\
    \midrule
    \multirow{2}{*}{\apidata} & \multirow{2}{*}{API Retrieval} & Java & Python & 140 & - \\
    &  & Python & Java & 140 & - \\
    \bottomrule
    \end{tabular}
    }
\end{table}

\textbf{Datasets.}
In consistent with our empirical study, we evaluate \ourname on the CodeNet \cite{codenet} and AVATAR \cite{avatar}. Additionally, we construct a new \apidata benchmark to assess the performance of the API sequence retriever, due to the lack of existing datasets. Table \ref{tab:datasets} provides detailed features and statistics for these datasets. 

The \apidata dataset is constructed from PolyHumanEval \cite{polyhumaneval}, a benchmark designed for cross-language code translation, including Python and Java. Specifically, we extract API call sequences from all Java and Python translations in PolyHumanEval, filtering out parallel translation pairs without API sequences. The resulting \apidata dataset contains 140 high-quality API sequence pairs in Python and Java, providing a robust foundation for evaluating API sequence retrieval tasks.

\textbf{Implementation Details.}
We utilize GPT-3.5-turbo as the backbone model, accessed via the official OpenAI API\footnote{\url{https://platform.openai.com/docs/api-reference}}, and employ its default parameter configuration for all translation tasks. For encoding API sequences and API mappings, we leverage the text-embedding-3-large model\footnote{\url{https://openai.com/index/new-embedding-models-and-api-updates}}, storing the resulting embeddings in a vector database with a flat index structure. All experiments are conducted on a Linux server running Ubuntu 23.10, equipped with two Nvidia GeForce RTX 4090 GPUs and CUDA version 12.0.

\subsection{Overall Effectiveness (RQ1)}

Table~\ref{tab:rq1} presents the performance comparison between \ourname and baseline methods. Overall, \ourname achieves superior performance across all translation scenarios on both datasets.

\begin{table}[t]
    \caption{Performance of Various Approaches for Code Translation}
    \centering
    \label{tab:rq1}
    \begin{tabular}{@{} l *{4}{>{\centering\arraybackslash}p{1.1cm}} @{}}
        \toprule
        \multirow{2}{*}{\textbf{Approach}} & \multicolumn{2}{c}{\textbf{CodeNet}} & \multicolumn{2}{c}{\textbf{AVATAR}} \\
        \cmidrule(lr){2-3} \cmidrule(lr){4-5}
         & \textbf{Py-to-Ja} & \textbf{Ja-to-Py} & \textbf{Py-to-Ja} & \textbf{Ja-to-Py} \\
        \midrule
        Direct translation & 82.0\% & 63.5\% & 49.6\% & 67.5\% \\
        Two-step translation & 86.5\% & 69.5\% & 53.6\% & 72.7\% \\
        EXP & 84.5\% & 65.5\% & 43.6\% & 60.8\% \\
        SpecTra & 87.5\% & 70.0\% & 54.0\% & 65.5\% \\
        \midrule
        \textbf{\ourname}   & \textbf{91.5\%} & \textbf{74.5\%} & \textbf{58.0\%} & \textbf{75.9\%} \\
        \bottomrule
    \end{tabular}
    \vspace{-5pt}
\end{table}

\begin{table*}[t!]
    \caption{Performance Across Different LLMs}
    \centering
    \renewcommand{\tablename}{Table}
    \label{tab:rq2}
    \begin{tabular}{@{} lc *{4}{>{\centering\arraybackslash}p{2.4cm}} @{}}
        \toprule
        \multirow{2}{*}{\textbf{Model}} & \multirow{2}{*}{\textbf{Approach}} & \multicolumn{2}{c}{\textbf{CodeNet}} & \multicolumn{2}{c}{\textbf{AVATAR}} \\
        \cmidrule(lr){3-4} \cmidrule(lr){5-6}
        &  & \textbf{Py-to-Ja} & \textbf{Ja-to-Py} & \textbf{Py-to-Ja} & \textbf{Ja-to-Py} \\
        \midrule
        \multirow{2}{*}{\textbf{StarCoder}}  & Direct translation & 48.5\%\ & 22.0\%\ & 22.0\%\ & 20.9\%\ \\
        & \ourname & 56.5\%(+16.5\%) & 34.0\%(+54.5\%) & 26.8\%(+21.8\%) & 27.3\%(+30.6\%) \\
        \midrule
        \multirow{2}{*}{\textbf{GPT-3.5-Turbo}}  & Direct translation & 82.0\%\ & 63.5\%\ & 49.6\%\ & 67.5\%\ \\
        & \ourname & 91.5\%(+11.6\%) & 74.5\%(+17.3\%) & 58.0\%(+17.0\%)\ & 75.9\%(+12.4\%) \\
        \midrule
        \multirow{2}{*}{\textbf{GPT-4o-mini}}  & Direct translation & 92.0\%\ & 64.0\%\ & 63.2\%\ & 64.7\%\ \\
        & \ourname & 95.5\%(+3.8\%)\ & 70.5\%(+10.2\%)\ & 69.2\%(+9.5\%)\ & 69.5\%(+7.4\%)\ \\
        \bottomrule
    \end{tabular}
    \vspace{-6pt}
\end{table*}

Compared to direct LLM translation and two-step translation, \ourname demonstrates significant improvements. Specifically, on the CodeNet dataset, \ourname outperforms these methods by an average CA score improvement of 10.3\% and 5.0\%, respectively. On the AVATAR dataset, the improvements are 8.4\% and 3.8\%, respectively. These gains can be attributed to \ourname's API knowledge augmented translation, thereby more effectively handling API calls. Additionally, the two-step translation method surpasses direct translation, benefiting from a chain-of-thought mechanism facilitated by API sequences.

Against EXP, \ourname achieves improvements of 7\% and 9\% on CodeNet for Python$\rightarrow$Java and Java$\rightarrow$Python, respectively, and 14.4\% and 15.1\% on AVATAR. While EXP performs moderately on CodeNet, it underperforms on AVATAR due to two key factors. First, EXP generates natural language explanations of the source program, which are effective for CodeNet's simpler tasks but produce noisy outputs for AVATAR's complex and diverse samples, degrading translation quality.
Second, EXP relies on translation templates similar to CodeGeeX \cite{codegeex}, which lack standardized references and require manual adjustments (e.g., ``$This\{\{\{ToFillTranslatedPythonCode\}\}\}$'' for Java$\rightarrow$Python translation). This generic approach introduces uncertainty,  potentially reducing code quality and consistency. 

Compared to SpecTra, \ourname demonstrates a performance boost of 4\% (Python$\rightarrow$Java) and 4.5\% (Java$\rightarrow$Python) on CodeNet, and 4\% and 10.4\%  on AVATAR. SpecTra exhibits a notable performance gap in the Java$\rightarrow$Python task on AVATAR, primarily due to its reliance on static templates that are less compatible with AVATAR's complexity, resulting in noisy specifications and degraded translation quality. Furthermore, SpecTra generates multiple references using real test cases, risking test case leakage. In contrast, our \ourname employs synthetic test cases derived directly from the original code, avoiding data leakage while potentially introducing minor noise.

\definecolor{my-blue}{rgb}{0.98, 0.98, 1.0} 
\begin{tcolorbox}[width=\linewidth, boxrule=0.8pt, left=2pt, right=2pt, top=2pt, bottom=2pt, colback=my-blue,]
\textbf{Answer to RQ1:} \ourname demonstrates significant performance improvements over baseline methods on both datasets, achieving a 4\% increase in CA score compared to the strong baseline, SpecTra.
\end{tcolorbox}

\subsection{Generalization (RQ2)}

To investigate the impact of the backbone model on \ourname's performance, we evaluate three widely-used LLMs: StarCoder, GPT-3.5-Turbo, and GPT-4o-mini. The results are presented in Table~\ref{tab:rq2}.

We can observe that \ourname significantly improves translation performance across all base models. StarCoder exhibits the most substantial improvements, with relative gains of 16.5\% and 54.5\% for Python$\rightarrow$Java and Java$\rightarrow$Python on CodeNet, and 21.8\% and 30.6\% on AVATAR. This is attributed to StarCoder's lower baseline performance, which offers greater improvement potential. GPT-3.5-Turbo also achieves notable gains, with improvements of 11.6\% and 17.0\% on CodeNet and AVATAR for Python$\rightarrow$Java. GPT-4o-mini, despite its strong baseline performance (92.0\% on CodeNet), still improves by 3.8\% and 9.5\% in the Python$\rightarrow$Java task on CodeNet and AVATAR, respectively.

These results highlight \ourname's generalizability, driven by its integration of three core techniques: \textit{target API sequence retrieval}, \textit{API sequence back-translation}, and \textit{API mapping}. By integrating diverse API knowledge, \ourname mitigates LLMs' limitations in domain-specific understanding, enhancing their ability to interpret code semantics and translate cross-language API logic. While StarCoder shows significant improvements, GPT models outperform due to two key factors: (1) their training on larger and more diverse datasets, enabling better semantic relationship capture, and (2) their advanced architectures and larger parameter scales, which fully leverage \ourname’s capabilities.

\definecolor{my-blue}{rgb}{0.98, 0.98, 1.0} 
\begin{tcolorbox}[width=\linewidth, boxrule=0.8pt, left=2pt, right=2pt, top=2pt, bottom=2pt, colback=my-blue,]
\textbf{Answer to RQ2:} \ourname consistently enhances the performance of various LLMs in code translation tasks, demonstrating its adaptability and effectiveness across different models.
\end{tcolorbox}

\subsection{Contribution of Each API Knowledge (RQ3)}

Table~\ref{tab:rq3} demonstrates the individual and combined contributions of each API knowledge component in \ourname to the overall translation performance. In general, all three components, namely \textit{API sequence retrieval}, \textit{API sequence back-translation}, and \textit{API mapping}, positively impact translation accuracy, with their combination yielding the best results.

\begin{table}[t!]
    \centering
    \caption{Contribution of API Knowledge Components}
    \renewcommand{\tablename}{Table}
    \label{tab:rq3}
    \resizebox{\columnwidth}{!}{%
    \begin{tabular}{lcccc}
        \toprule
        \multirow{2}{*}{\textbf{API Knowledge Component}} & \multicolumn{2}{c}{\textbf{CodeNet}} & \multicolumn{2}{c}{\textbf{AVATAR}} \\
        \cmidrule(lr){2-3} \cmidrule(lr){4-5}
         & \textbf{Py-to-Ja} & \textbf{Ja-to-Py} & \textbf{Py-to-Ja} & \textbf{Ja-to-Py} \\
        \midrule
        None & 82.0\% & 63.5\% & 49.6\% & 67.5\% \\
        API Sequence & 85.5\% & 71.5\% & 54.8\% & 73.9\% \\
        API Mapping & 86.0\% & 68.5\% & 54.0\% & 72.3\% \\
        API Sequence + API Mapping & 88.5\% & 72.5\% & 55.2\% & 74.8\% \\
        API Sequence + Back-translation & 90.5\% & 74.0\% & 55.6\% & \textbf{75.9\%} \\
        \textbf{All (\ourname)} & \textbf{91.5\%} & \textbf{74.5\%} & \textbf{58.0\%} & \textbf{75.9\%} \\
        \bottomrule
    \end{tabular}     
    }
    \vspace{-6pt}
\end{table}

\textit{API sequence retrieval} alone delivers the most substantial improvement in translation accuracy, underscoring its efficacy in guiding the translation process. In contrast, while \textit{API mapping} also enhances performance, its impact is less pronounced compared to \textit{API sequence retrieval}. Combining these two techniques yields further improvements, as the recommended target API sequences offer a clearer translation objective, while API mappings establish direct correspondences between source and target APIs. These two granularities of API knowledge complement each other: one emphasizes API structures, while the other focuses on individual APIs.

Furthermore, integrating \textit{API sequence back-translation} amplifies this effect, particularly when paired with \textit{API sequence retrieval}, as it enhances the model's understanding of cross-language API logic. By combining all three components, \ourname achieves the highest accuracy, effectively addressing the limitations of individual knowledge sources and maximizing overall performance.

Notably, in the Java$\rightarrow$Python task on the AVATAR dataset, \textit{API mapping} does not yield additional improvements. This may be attributed to redundancy between recommended target API sequences and API mappings, as the essential knowledge is already sufficiently captured by the API sequences.

\definecolor{my-blue}{rgb}{0.98, 0.98, 1.0} 
\begin{tcolorbox}[width=\linewidth, boxrule=0.8pt, left=2pt, right=2pt, top=2pt, bottom=2pt, colback=my-blue,]
\textbf{Answer to RQ3:} 
\textit{API sequence retrieval}, \textit{API sequence back-translation}, and \textit{API mapping} each contribute to enhancing the code translation performance of \ourname, with their combined integration achieving the optimal results.
\end{tcolorbox}

\subsection{Effectiveness of API Knowledge Retriever (RQ4)}

Our method leverages the text-embedding-3-large model to implement the API knowledge retriever, facilitating the retrieval of both API sequences and API mappings. We evaluate its performance against one text retrieval algorithm (i.e., BM25) and four code search models (i.e., CodeBERT, GraphCodeBERT, UniXcoder, and StarEncoder) on the \apidata dataset.

\begin{itemize}
    \item \textbf{BM25} \cite{bm25} is a widely-used text retrieval algorithm that assesses the relevance between search terms and documents based on term frequency and document length. In this work, we apply the standard Okapi-BM25 algorithm with parameters set to $k_1 = 1.5$ and $b = 0.75$.
    
    \item \textbf{CodeBERT} \cite{codebert} is the first pre-trained model designed for multiple programming languages with an encoder-only architecture. It encodes API calls directly, followed by similarity computation. Since CodeBERT lacks pretraining specific to API knowledge, we fine-tune it using contrastive learning on the CodeSearchNet  \cite{codesearchnet} dataset.

    \item \textbf{GraphCodeBERT} \cite{graphcodebert} extends CodeBERT by incorporating graph-based structural information to enhance code understanding and generation tasks. Similarly, we fine-tune it using contrastive learning on CodeSearchNet.
    
    \item \textbf{UniXCoder} \cite{unixcoder} is a unified pre-trained model supporting three architectures (i.e., Encoder-Decoder, Encoder-only, and Decoder-only). It introduces two pre-training tasks, namely Multi-Modal Contrastive Learning and Cross-Modal Generation, to acquire semantic embeddings of code. For API knowledge retrieval, UniXcoder operates in encoder-only mode to encode APIs.

    \item \textbf{StarEncoder} \cite{starcoder} is an encoder-only pre-trained model developed alongside StarCoder. It does not provide a default sentence embedding method. We evaluate two methods: average pooling and using the embedding corresponding to the special token at the input's start position, selecting the optimal one for encoding APIs.
   
\end{itemize}

The results are presented in Table \ref{tab:rq4}. The text-embedding-3-large model significantly outperforms other methods, achieving accuracy rates of 97.85\% and 98.57\% in Python$\rightarrow$Java and Java$\rightarrow$Python, respectively. This demonstrates the model's exceptional accuracy and stability in handling API knowledge retrieval tasks.

\begin{table}[t]
    \caption{Accuracy of Various API knowledge Retrievers}
    \centering
    \label{tab:rq4}
    \begin{tabular}{lcc}
        \toprule
        \textbf{API Knowledge Retriever} & Py-to-Ja & Ja-to-Py \\
        \midrule
        BM25 & 34.29\% & 35.71\% \\
        CodeBERT & 13.57\% & 17.85\% \\
        GraphCodeBERT & 54.28\% & 60.00\% \\
        UniXCoder & 71.42\% & 72.85\% \\
        StarEncoder & 39.28\% & 37.14\% \\
        Text-embedding-3-large & \textbf{97.85}\% & \textbf{98.57\%} \\
        \bottomrule
    \end{tabular}
    \vspace{-8pt}
\end{table}

BM25 surpasses CodeBERT and performs comparably to StarEncoder, yet remains significantly inferior to other embedding-based methods. This highlights the limitations of keyword-based approaches relative to semantic embedding techniques for API knowledge retrieval.

UniXcoder exhibits strong performance, likely attributable to its inclusion of API-related tasks during pretraining. GraphCodeBERT also performs competitively, achieving robust results after contrastive learning fine-tuning despite lacking API-specific pretraining. StarEncoder delivers moderate performance, potentially hindered by the absence of a default sentence embedding method. CodeBERT underperforms in both tasks, possibly due to its limited natural language understanding, which may compromise contrastive learning effectiveness.

\definecolor{my-blue}{rgb}{0.98, 0.98, 1.0} 
\begin{tcolorbox}[width=\linewidth, boxrule=0.8pt, left=2pt, right=2pt, top=2pt, bottom=2pt, colback=my-blue,]
\textbf{Answer to RQ4:} Embedding-based retrievers consistently outperform keyword-based approaches in API knowledge retrieval. Among the evaluated methods, OpenAI's embedding model achieves the highest performance, demonstrating superior accuracy and stability.
\end{tcolorbox}

\subsection{Impact of Parameter Settings (RQ5)}
\begin{figure}[t!]
\centerline{\includegraphics[width=0.5\textwidth, trim=0 10 0 50, clip]{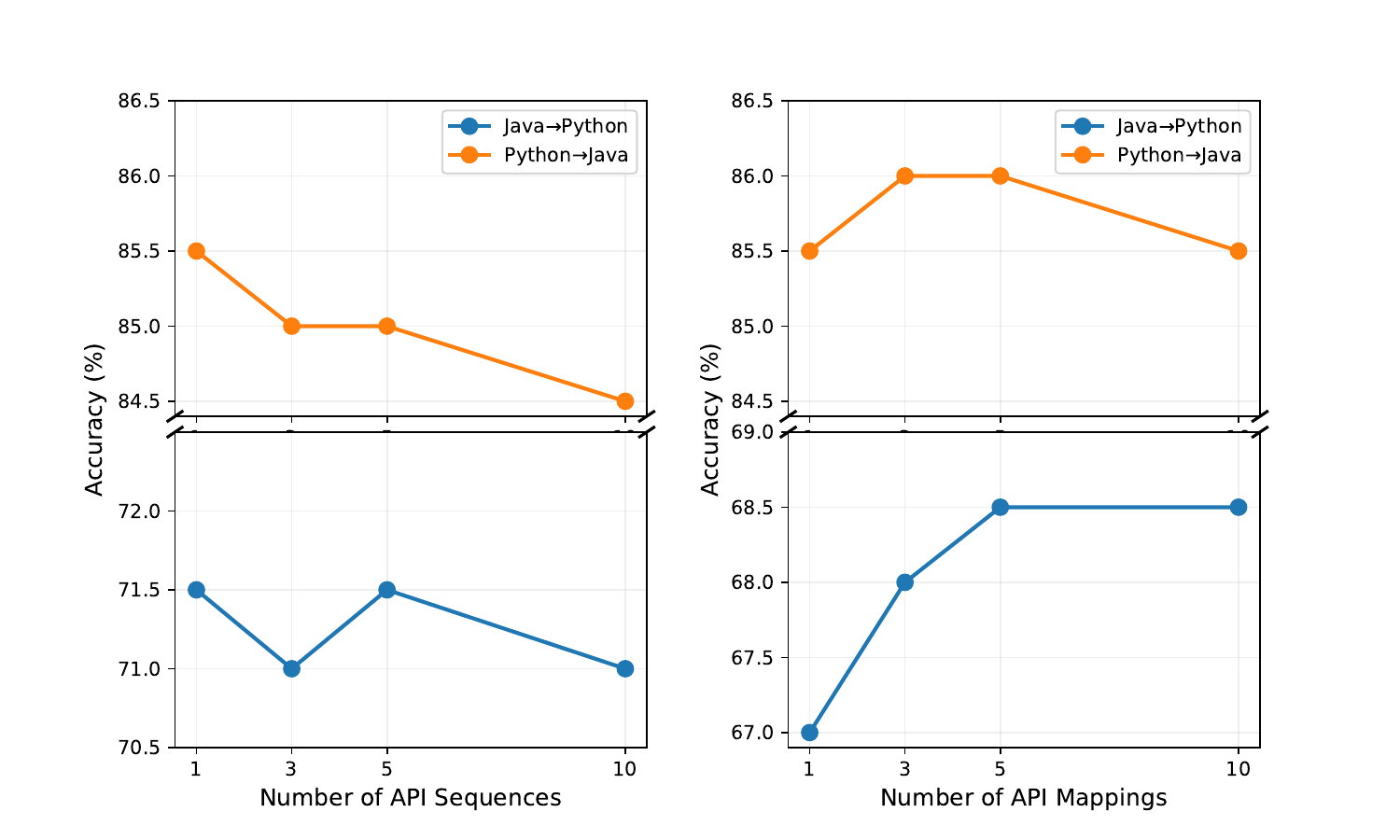}}
    \caption{Performance with Different Parameter Settings} 
    \label{fig:rq5}
    \vspace{-8pt}
\end{figure}

The performance of retrieval-augmented generation is highly sensitive to the number of retrieved samples. Therefore, we conduct an ablation study on CodeNet to investigate the impact of the number of retrieved API sequences and API mappings, evaluating \ourname in two configurations: (1) using only API sequences and (2) using only API mappings. The experimental results are presented in Figure \ref{fig:rq5}.

\textbf{Number of API Sequences}. For Python$\rightarrow$Java translation, performance peaks at 85.5\% when k=1, but slightly declines to 84.5\% as k increases. This indicates that including additional sequence samples introduces low-quality candidates, which can negatively impact translation quality. In contrast, for Java$\rightarrow$Python translation, performance remains stable, ranging from 71.0\% to 71.5\% as k increases, suggesting that the number of candidate sequences has minimal effect due to the concentrated distribution of target sequences.

\textbf{Number of API Mappings}. For Python$\rightarrow$Java translation, performance reaches 85.5\% at n=1, improving marginally  to 86.0\% as n increases to 3 and 5. However, at n=10, performance drops back to 85.5\%, likely due to the introduction of redundant information from excessive mappings. For Java$\rightarrow$Python translation, performance gradually improves from 67.0\% to 68.5\%, demonstrating greater sensitivity to the number of API mappings. 


\definecolor{my-blue}{rgb}{0.98, 0.98, 1.0} 
\begin{tcolorbox}[width=\linewidth, boxrule=0.8pt, left=2pt, right=2pt, top=2pt, bottom=2pt, colback=my-blue,]
\textbf{Answer to RQ5}: The performance of \ourname is influenced by the number of retrieved API sequences and API mappings. Optimal results are achieved when the number of API mappings is set to 5 and the number of API sequences is set to 1. 
\end{tcolorbox}

\section{Discussions}

\subsection{Future Directions}

We suggest three possible future research directions for API knowledge augmented code translation with LLMs.

First, API knowledge can be expanded beyond traditional sources such as API documentation and existing code bases. Future research could explore integrating diverse knowledge sources into prompt design and knowledge augmentation frameworks. For syntax-related enhancements, incorporating comprehensive information about third-party libraries, such as class descriptions, class hierarchies, and inter-class relationships, could prove beneficial. For semantic understanding, leveraging descriptive and explanatory content from platforms like Stack Overflow may provide valuable insights. Additionally, domain-specific knowledge, such as translation rules and code templates, could further enrich the translation process.

Second, our results indicate that a dedicated retriever can effectively retrieve API knowledge from pre-built databases. Future work could focus on developing more advanced knowledge retrieval mechanisms. For instance, researchers could design a question-answering system or search engine capable of dynamically retrieving factual knowledge during the code translation process.

Third, our findings highlight the effectiveness of incorporating API knowledge through prompt engineering during the inference stage. Future research could investigate integrating API knowledge at earlier stages of LLM development, such as during pre-training or fine-tuning. Instead of training language models solely on plain code, augmenting training data with API knowledge could enhance the model's ability to understand and generate code with greater accuracy and contextual relevance.

\subsection{Threats to Validity and Limitations}

Our experiments are conducted on three LLMs, which might not represent all the latest advancements in large-scale models. 
Furthermore, our method focuses on function-level code translation, which does not address the repository-level issues at real-world scenarios.
Regarding the evaluation metrics, we have utilized Computational Accuracy (CA). Although CA is widely used in code intelligence research, they may not fully represent human experience.

\section{Related Work}
\label{sec:related}

\subsection{Code Translation}
Traditional rule-based code translation, like Stratego/XT~\cite{rule} and c2rust~\cite{c2rust}, relies on predefined rules. However, such approaches require human experts to invest significant time and manual efforts in crafting rules, and the translated target program often suffers from poor readability and usability.

To address these issues, a series of supervised learning methods have been proposed to improve translation effectiveness \cite{codebert, codet5}. These methods train models on large amounts of parallel data, allowing models to learn the translation patterns and mappings between different languages. However, their performance and generalizability are restricted by the scarcity and quality of parallel data. 
Unsupervised learning methods alleviate such limitations. TransCoder~\cite{transcoder} leverages back-translation to train models without parallel data. TransCoder-ST\cite{transcoder-st} and TransCoder-IR\cite{transcoder-ir} thereafter leverage unit tests and compiler representations for further optimization. 
However, they still require specialized training for each programming language pair and thus demand substantial computation resources for generalization.

In recent years, code translation based on LLMs has become the mainstream approach due to their powerful cross-lingual understanding and representation, as well as their in-context learning ability. 
Jiao et al.~\cite{gtranseval} proposed a 4-type taxonomy for code translation and evaluated LLMs' performance on each type of translation task. 
Pan et al.~\cite{lostintranslation} generalized 15 bug categories from LLMs' unsuccessful code translations. 
Yang et al.~\cite{unitrans} proposed UniTrans, a code translation framework leveraging LLMs by generating test cases and repairing.
Tang et al.~\cite{exp} proposed a translate-after-explain framework, where the model first generates an explanation of the source program before translating it. 
Bhattarai et al.~\cite{bhattarai2024} improved code translation in LLMs by employing few-shot learning through retrieval-augmented generation. 
SpecTra~\cite{spectra} explores various types of specifications extracted from programs to enhance the code translation capabilities of LLMs. 
TransAgent~\cite{transagent} presents an LLM-based multi-agent system for code translation.
However, LLM-based code translation methods still face limitations when handling cross-lingual API translation, often resulting in inaccurate or incomplete outputs. Our approach addresses this issue by integrating multi-source API knowledge into LLMs.

\subsection{API Knowledge Augmented Code Generation}

In the broader domain of code generation, API knowledge augmentation techniques have been extensively studied, with numerous approaches proposed to tackle related challenges and exploit emerging opportunities. For example, Zan et al.~\cite{zan2022language} introduced a retrieval-then-coding framework in which an API retriever first identifies relevant APIs, followed by an API coder that generates code utilizing these APIs. Lin et al.~\cite{domcoder} developed the DomCoder method, which employs a GPT-based API recommender to generate API knowledge prompts for LLMs and formulates code generation as a chain-of-thought process, integrating APIs at each intermediate coding step. 
Ma et al.~\cite{Ma_2024} proposed a decomposed retrieval method, incorporating an inter-task LLM reranker to suggest k APIs for code generation. Wang et al.~\cite{exploracoder} proposed ExploraCoder to guide the LLM to iteratively generate several experimental API invocations for each simple subtask, prior to final code generation.

Building upon these advancements in code generation, we aim to elucidate the unique challenges and opportunities for API knowledge integration specifically within the context of code translation. We propose three novel retrieval-augmented translation techniques that leverage multi-source API knowledge to deepen the model's understanding of API usage patterns, thereby facilitate more precise and coherent code translations.

\section{Conclusion}

In this paper, we introduce \ourname, a novel code translation method based on LLMs that integrates multi-source API knowledge to mitigate API mistranslation errors. Our method designs three retrieval-augmented techniques, namely \textit{API sequence retrieval}, \textit{API sequence back-translation}, and \textit{API mapping}, to enhance the LLMs' comprehension of both the structural logic of API sequences and the usage patterns of individual APIs. 
Our experimental results demonstrate that \ourname significantly outperforms existing methods in terms of computational accuracy and exhibits robust stability across various datasets and translation directions. 
We hope our work will inspire future research aimed at advancing the code translation capabilities of LLMs in real-world applications.


\section*{Acknowledgment}
This research is supported by National Key R\&D Program of China (Grant No. 2023YFB4503802) and National Natural Science Foundation of China (Grant No. 62102244, 62232003).

\balance
\bibliographystyle{IEEEtran.bst}
\bibliography{ref.bib}

\end{document}